\documentstyle{article}

\def\be{\begin{equation}}
\def\ee{\end{equation}}
\def\bea{\begin{eqnarray}}
\def\eea{\end{eqnarray}}
\begin{document}

\pagestyle{empty}
\vskip-10pt
\hfill {\tt hep-th/0104172}

\begin{center}
\vskip 5truecm
{\LARGE \bf
A short representation of the six-dimensional $(2, 0)$ algebra}\\ 
\vskip 2truecm
{\large \bf Andreas Gustavsson and M{\aa}ns Henningson}\\
\vskip 1truecm
{\it Institute of Theoretical  Physics, Chalmers University of Technology\\
S-412 96 G\"{o}teborg, Sweden}\\
\vskip 5truemm
{\tt f93angu@fy.chalmers.se, mans@fy.chalmers.se}
\end{center}
\vskip 2truecm
\noindent{\bf Abstract:}
We construct a BPS-saturated representation of the six-dimensional $(2, 0)$ algebra with a certain non-zero value of the `central' charge. This representation is naturally carried by strings with internal degrees of freedom rather than by point particles. Upon compactification on a circle, it reduces to a massive vector multiplet in five dimensions. We also construct quantum fields out of the creation and annihilation operators of the states of this representation, and show how they give rise to a conserved two-form current that can be coupled to a tensor multiplet. We hope that these results may be relevant for understanding the degrees of freedom associated with strings in interacting $(2, 0)$ theories.
\newpage
\pagestyle{plain}

\section{Introduction}
Our understanding of the $(2, 0)$ superconformal theories in six dimensions \cite{Witten} is still very incomplete, and we do not even know what are the correct degrees of freedom necessary to give a proper formulation of them. An important clue to their structure is given by considering compactification on a small circle, upon which the $(2, 0)$ theories reduce to five-dimensional maximally supersymmetric Yang-Mills theories with a simply laced gauge group. (See e.g. \cite{Seiberg}.) At a generic point in the moduli space, the gauge group $G$ of such a five-dimensional super Yang-Mills theory is spontaneously broken to $U (1)^r$, where $r$ is the rank of $G$. The degrees of freedom of the theory are then given by a massless vector multiplet for each Cartan generator of $G$, and a massive vector multiplet for each root generator of $G$. The mass of the latter multiplets of course arises through the Higgs mechanism, so at the origin of the moduli space, where the full symmetry group $G$ is restored, we have a massless vector multiplet for each generator of $G$. We will now try to understand the six dimensional origin of these facts. The massless vector multiplets associated with the Cartan generators of $G$ arise from so called tensor multiplets of the $(2, 0)$ algebra, each comprising a chiral two-form $X$ (i.e. with self-dual field strength $d X = * d X$), an $Sp (4)$ symplectic Majorana-Weyl fermion $\Psi$ and an $SO (5)$ vector $\Phi^a$ of scalars. ($Sp (4) \simeq SO (5)$ is the $R$-symmetry of the $(2, 0)$ algebra.) The purpose of this letter is to construct a new representation of the $(2, 0)$ algebra, that similarly might give rise to the massive vector multiplets associated with the root generators of $G$.

One might first think that the answer should be some kind of massive tensor multiplet, the mass of which arises from a generalized Higgs mechanism applied to an ordinary massless tensor multiplet. One would thus try to construct a kind of non-abelian theory with a set of tensor multiplets transforming in the adjoint representation of the gauge group. There have indeed been attempts in this direction, but no convincing solution has emerged so far. In retrospect, one can understand the problems with this approach: The various fields in the tensor multiplets (chiral tensors, spinors and scalars) associated with the Cartan algebra generators of $G$ describe point particles with different internal degrees of freedom. The degrees of freedom associated with the root generators of $G$ should couple to these fields, and in particular to the chiral tensors. But such tensors naturally couple to strings rather than to point particles. We should therefore look for a suitable multiplet that can describe strings with different internal degrees of freedom. Upon compactification on a circle, such strings may wind around the compact direction, thus giving rise to point particles with different internal degrees of freedom in five dimensions. 

The next question is what the correct internal degrees of freedom of such strings are in order for them to give rise to the desired particles in five dimensions. The internal degrees of freedom of a massive point particle can be described by going to the rest frame of the particle and considering its behaviour under the `little group' of purely spatial rotations. Similarly, the internal degrees of freedom of a string can be described by going to a frame with respect to which the string is at rest pointing in a specified direction and considering its behaviour under the `little group' of spatial rotations transverse to this direction. Strings that transform as a vector, spinors and scalars under this little group will give rise to massive particles in five dimensions that also transform as a vector, spinors and scalars, i.e. to a massive vector multiplet. This is to be contrasted with the massless case, where a particle transforming as a chiral tensor in six dimensions gives rise to a particle transforming as a vector in five dimensions. 

We are thus led to considering a theory with a tensor multiplet associated with each Cartan generator of a simply laced Lie algebra $G$, and one of the `vector-string' multiplets described above associated with each root generator of $G$. The six-dimensional theory would thus not have $G$ as a symmetry group; such a symmetry would only appear after compactification to five dimensions (and after the spontaneous symmetry breaking is turned off). We do not know whether such a theory can indeed be constructed, but we would like to point out the following encouraging difference compared to six-dimensional Yang-Mills theory (which we do {\it not} think is a consistent quantum theory): In six dimensions, the coupling constant for a vector field has mass dimension $-1$, and the theory is thus non-renormalizable by power counting. By contrast, the coupling constant for a two-form in six dimensions is dimensionless, so here we might have a better hope for a renormalizable theory. Furthermore, if the two-form is chiral, the coupling constant has to take a particular `self-dual' value \cite{Henningson-Nilsson-Salomonson}, indicating that the theory is a fixed point of the renormalization group as expected. We would also like to point out, that even if this approach might seem unfamiliar, it is actually rather conservative: On very general grounds, we know that the Hilbert space of a quantum theory decomposes into unitary representations of the symmetries of the theory. The study of representations of the $(2, 0)$ algebra should therefore a priori be relevant for understanding the $(2, 0)$ theories.

The multiplet that we will be considering is a `short' or BPS-saturated representation of the $(2, 0)$ algebra in the following sense: The supercharges of this algebra are Weyl spinors that fulfil a symplectic Majorana condition. The anti-commutator of two such supercharges is a linear combination of the momentum operator and certain `central' charges. We will consider a particular case of such a charge that arises in the supersymmetry algebra of a tensor multiplet in the presence of a string that couples to the chiral two-form and with a non-trivial vacuum expectation value for the scalars. Given the eigenvalue of this charge operator, unitarity of the representation imposes a lower bound for the eigenvalue of the mass operator. When this bound is saturated, half of the supercharges annihilate the state, and consequently the representation is smaller than a generic representation. Being a `string' representation, it is still infinitely  bigger than a `particle' representation such as the tensor multiplet, though.

The formalism with a little group is natural from a representation theoretic point of view and suffices for describing a single excitation of the vacuum. It is however rather inconvenient for constructing interacting theories. This can be remedied by introducing quantum fields that transform covariantly under the full Lorentz and $R$-symmetry groups. Since our excitations are string like rather than particle like, these fields will depend not only on space-time position but also on the direction of the string, both in ordinary space and the $R$-symmetry space. As a first step towards an interacting theory, we will briefly describe how to construct a conserved two-form current from these fields, that can be coupled to the chiral two-form in a tensor multiplet.

\section{The $(2, 0)$ algebra}
The bosonic part of the $(2, 0)$ algebra is spanned by the momentum operators $P_\mu$ for $\mu = 0, 1, \ldots, 5$, the $SO (5, 1)$ Lorentz generators $J_{\mu \nu} = - J_{\nu \mu}$, the $SO (5) \simeq Sp (4)$ $R$-symmetry generators $T^{a b} = - T^{b a}$ for $a, b = 1, \ldots 5$ and the `central' charges $Z_\mu^a$ and $W_{\mu \nu \rho}^{a b}$ \cite{Howe-Lambert-West}. The latter charge is totally antisymmetric in both sets of indices and is furthermore self-dual as a space-time three-form. (These charges are not truly central in the algebra, since they behave non-trivially under Lorentz and $R$-symmetry transformations.) The fermionic generators are denoted as $Q_\alpha^i$, where $\alpha$ is an $SO (1, 5)$ spinor index and $i$ is an $SO (5)$ spinor index (or equivalently an $Sp (4)$ fundamental index). They are Weyl spinors, i.e. $\left(\Gamma^7 \right)_\alpha{}^\beta Q_\beta^i = Q_\alpha^i$, where $\Gamma^7$ is the $SO(5, 1)$ chirality operator, and also fulfil the symplectic Majorana condition $\left(Q_\alpha^i\right)^\dagger = \Omega_{i j} \left(\Gamma^0 \right)^{\alpha \beta} Q_\beta^j$, where $\Omega_{i j}$ is the anti-symmetric $Sp (4)$ invariant. ($R$-symmetry and Lorentz spinor indices are raised and lowered with $\Omega_{i j}$, the charge conjugation matrix $C_{\alpha \beta}$, and their inverses $\Omega^{i j}$ and $C^{\alpha \beta}$ respectively.) The anti-commutator of two supercharges is
\be
\left\{Q_\alpha^i, Q_\beta^j \right\} = \Omega^{i j} \left(\Gamma^\mu \right)_{\alpha \beta} P_\mu + \left(\Gamma^\mu \right)_{\alpha \beta} \left(\gamma_a \right)^{i j} Z_\mu^a + \left(\Gamma^{\mu \nu \rho} \right)_{\alpha \beta} \left(\gamma_{a b} \right)^{i j} W_{\mu \nu \rho}^{a b} . \label{algebra}
\ee

The operators $P_\mu$, $Z_\mu^a$ and $W_{\mu \nu \rho}^{a b}$ all commute with each other, and one may therefore introduce a basis of simultaneous eigenvectors of them in the representation space. A priori, the corresponding eigenvalues $p_\mu$, $z_\mu^a$ and $w_{\mu \nu \rho}^{a b}$ are arbitrary, but in this letter, we will content ourselves with analyzing the special case where the following constraints are imposed:
\bea
w_{\mu \nu \rho}^{a b} & = & 0 \cr
p^\mu p_{\mu} & = & - m^2 \cr
{\rm rank} \; z_\mu^a & = & 1 \cr
p^\mu z_\mu^a & = & 0 
\eea 
for some real constant $m$. The third equation means that the matrix $z_\mu^a$ can be written as $z_\mu^a = v_\mu \phi^a$ for some $SO (5, 1)$ vector $v_\mu$ and some $SO (5)$ vector $\phi^a$. According to the last equation, $v_\mu$ then obeys $p^\mu v_\mu = 0$. We have imposed these conditions because they arise in a physically interesting situation, as we will now describe. This can be most easily seen by going to the rest frame so that $p_\mu$ is purely time-like. We then get that $v_\mu$ is a purely space-like vector, which we can take to lie on the unit four-sphere, i.e. $v^\mu v_\mu = 1$, by absorbing its length in the vector $\phi^a$. A charge $z_\mu^a$ fulfilling these requirements arises in the supersymmetry algebra of a tensor multiplet in the presence of a static string coupled to the chiral tensor field and pointing in the $v_\mu$ direction, and with non-vanishing expectation values $\left< \Phi^a \right> = \phi^a$ of the scalar fields. The general solution to the equations $p_\mu p^\mu = - m^2$, $p_\mu v^\mu = 0$, and $v_\mu v^\mu = 1$ can be parametrized as
\bea
p_\mu & = & \left(\sqrt{{\bf p}^2 + m^2}, {\bf p} \right) \cr
v_\mu & = & \left(\frac{{\bf p \cdot \hat{u}}}{\sqrt{{\bf p}^2 + m^2 - ({\bf p \cdot \hat{u}})^2}},  \bf{\hat{u}} \sqrt{\frac{{\bf p}^2 + m^2}{{\bf p}^2 + m^2 - ({\bf p \cdot \hat{u}})^2}} \right) , \label{parametrization}
\eea
where the only constraint on the spatial vectors ${\bf p}$ and ${\bf \hat{u}}$ is that ${\bf \hat{u} \cdot \hat{u}} = 1$. A short calculation then gives the invariant `on shell' volume element as
\be
d^6 p d^6 v \delta (p_\mu p^\mu + m^2) \delta (p_\mu v^\mu) \delta (v_\mu v^\mu - 1) = d^5 {\bf p} d^4 {\bf \hat{u}} \frac{({\bf p}^2 + m^2)^{3/2}}{4 ({\bf p}^2 + m^2 - ({\bf p \cdot \hat{u}})^2)^{5/2}} . \label{volume_element}
\ee

\section{The representations}
To construct the representations of the $(2, 0)$ algebra, we will use the method of induced representations as is familiar from the study of the Poincar\'e algebra. A lucid presentation can be found in \cite{Weinberg}, and we will mostly use an analogous notation.

The representation that we will describe has a basis of simultaneous eigenstates $\left| p, v, \phi, \sigma \right>$ of the operators $P_\mu$ and $Z_\mu^a$, i.e.
\bea
P_\mu \left| p, v, \phi, \sigma \right> & = & p_\mu \left| p, v, \phi, \sigma \right> \cr
Z_\mu^a \left| p, v, \phi, \sigma \right> & = & v_\mu \phi^a \left| p, v, \phi, \sigma \right> .
\eea
The label $\sigma$ takes its values in some finite set and describes all other quantum numbers of the state. To describe the action of the other operators of $(2, 0)$ algebra on these states, we note that any $p_\mu$, $v_\mu$ and $\phi^a$ such that $p^\mu p_\mu = - m^2$, $p^\mu v_\mu = 0$, $v^\mu v_\mu = 1$, and $\phi^a \phi_a = \phi^2$ can be obtained by acting with some standard Lorentz and $R$-symmetry transformations $L (p, v)$ and $R (\phi)$ on the standard vectors
\bea
p_\mu^o & = & (m, 0, 0, 0, 0, 0) \cr
v_\mu^o & = & (0, 0, 0, 0, 0, 1) \cr
\phi_a^o & = & (0, 0, 0, 0, \phi) . \label{frame}
\eea
We then define the states $\left| p, v, \phi, \sigma \right>$ by the equation
\be
\left| p, v, \phi, \sigma \right> = N (p, v) L (p, v) R (\phi) \left| p^o, v^o, \phi^o, \sigma \right> ,
\ee
where $N (p, v)$ is a numerical normalization factor. This equation thus relates the $\sigma$ label for different values of $p_\mu$, $v_\mu$ and $\phi^a$. If we choose
\be
N (p, v) = \frac{m ({\bf p}^2 + m^2)^{3/4}}{({\bf p}^2 + m^2 - ({\bf p \cdot \hat{u}})^2)^{5/4}} ,
\ee
we find that the Hilbert space inner product is given by
\be
\left< p^\prime, v^\prime, \phi^\prime, \sigma^\prime \right| \left| p, v, \phi, \sigma \right> = \delta^5 ({\bf p} - {\bf p}^\prime) \delta^4 ({\bf \hat{u}} - {\bf \hat{u}}^\prime) \delta^5 (\phi - \phi^\prime) \delta_{\sigma \sigma^\prime} ,
\ee
where we have used the parametrization (\ref{parametrization}) and the result (\ref{volume_element}).

Acting on the state $\left| p, v, \phi, \sigma \right>$ with arbitrary Lorentz and $R$-symmetry transformations $\Lambda$ and ${\cal R}$, we find that
\bea
\Lambda \left| p, v, \phi, \sigma \right> & = & N (p, v) L (\Lambda p, \Lambda v) W (\Lambda, p, v) \left| p^o, v^o, \phi, \sigma \right> \cr
{\cal R} \left| p, v, \phi, \sigma \right> & = & R ({\cal R} \phi) Y ({\cal R}, \phi) \left| p, v, \phi^o, \sigma \right> ,
\eea 
where
\bea
W (\Lambda, p, v) & = & L^{-1} (\Lambda p, \Lambda v) \Lambda L (p, v) \cr
Y ({\cal R}, \phi) & = & R^{-1} ({\cal R} \phi) {\cal R} R (\phi) . 
\eea
The Lorentz transformation $W (\Lambda, p, v)$ belongs to the little subgroup $SO (4)_L$, that leaves the standard vectors $p^o$ and $v^o$ invariant. Similarly, the $R$-symmetry transformation $Y ({\cal R}, \phi)$ belongs to the little subgroup $SO (4)_R$, that leaves the standard vector $\phi^o$ invariant. It follows that these transformations must act on the $\sigma$ label only, i.e. they are given by some matrices $W_{\sigma \sigma^\prime} (\Lambda, p, v)$ and $Y_{\sigma \sigma^\prime} ({\cal R}, \phi)$ that represent $SO (4)_L$ and $SO (4)_R$ respectively. The transformation laws can then be written as
\bea
\Lambda \left| p, v, \phi, \sigma \right> & = & \frac{N (p, v)}{N (\Lambda p, \Lambda v)} \sum_{\sigma^\prime} W_{\sigma \sigma^\prime} (\Lambda, p, v) \left| \Lambda p, \Lambda v, \phi, \sigma^\prime \right> \cr
{\cal R} \left| p, v, \phi, \sigma \right> & = & \sum_{\sigma^\prime} Y_{\sigma \sigma^\prime} ({\cal R}, \phi) \left| p, v, {\cal R} \phi, \sigma^\prime \right> .
\eea
An irreducible multiplet of the bosonic part of the $(2, 0)$ algebra is thus completely specified by giving a representation of the little group $SO (4)_L \times SO (4)_R$. The fermionic operators couple several such multiplets together to form an irreducible super multiplet of the complete $(2, 0)$ algebra.

\section{BPS-saturated representations}
In our standard frame (\ref{frame}), the supersymmetry algebra (\ref{algebra}) reads
\be
\left\{ Q_\alpha^i, (Q_\beta^j)^\dagger \right\} = \delta_\alpha^\beta \delta_j^i m + \left(\Gamma^0 \Gamma^5 \right)_\alpha{}^\beta \left(\gamma_5 \right)^i{}_j \phi ,
\ee
where we have used the symplectic Majorana condition on $Q_\alpha^i$. Unitarity of the representation thus requires that 
\be
m \geq \phi .
\ee
We will be interested in the case of a BPS-representation where this inequality is saturated, i.e. $m = \phi$. This means that
\be
\left\{ Q_\alpha^i, (Q_\beta^j)^\dagger \right\} = 2 \Pi_{\alpha j}^{\beta i} \phi ,
\ee
where
\be
\Pi_{\alpha j}^{\beta i} = \frac{1}{2} \left(\delta_\alpha^\beta \delta_j^i + (\Gamma^0 \Gamma^5)_\alpha{}^\beta (\gamma_5)^i{}_j \right) .
\ee
It is convenient to use that 
\be
SO (4)_L \times SO (4)_R \simeq SU(2) \times SU (2) \times SU (2) \times SU (2) . \label{group}
\ee
The supercharge $Q_\alpha^i$ transforms in the Dirac spinor representation under both the Lorentz little group $SO (4)_L$ and the $R$-symmetry little group $SO (4)_R$, i.e. as
\be
(\frac{1}{2}, 0, \frac{1}{2}, 0) \oplus (0, \frac{1}{2}, \frac{1}{2}, 0) \oplus (\frac{1}{2}, 0, 0, \frac{1}{2}) \oplus (0, \frac{1}{2}, 0, \frac{1}{2}) ,
\ee
where we have labelled $SU (2)$ representations by their spin $j$ in the usual way. The operator $\Pi_{\alpha j}^{\beta i}$ is a projector onto the part of $Q_\alpha^i$ that transforms as a Weyl spinor under both $SO (4)_L$ and $SO (4)_R$ or as an anti-Weyl spinor under both groups, i.e. as
\be
(\frac{1}{2}, 0, \frac{1}{2}, 0) \oplus (0, \frac{1}{2}, 0, \frac{1}{2}) . 
\ee
(To see this, note that $\Gamma^0 \Gamma^5 = \Gamma^1 \Gamma^2 \Gamma^3 \Gamma^4$ and $\gamma_5 = \gamma_1 \gamma_2 \gamma_3 \gamma_4$ are the $SO (4)_L$ and $SO (4)_R$ chirality operators respectively. Here we have used that $\Gamma^0 \ldots \Gamma^5 = 1$ when acting on an $SO (5, 1)$ Weyl spinor.) The part of $Q_\alpha^i$ that is projected out should be put to zero when constructing the representation. We label the surviving components of $Q_\alpha^i$ by their $J^z$ eigenvalue $m$ for the four different $SU (2)$ groups in the usual way. In this way we get the components
\be
(- \frac{1}{2}, 0, -\frac{1}{2}, 0) \;\;,\;\; (- \frac{1}{2}, 0, + \frac{1}{2}, 0) \;\;,\;\; (0, -\frac{1}{2}, 0, - \frac{1}{2}) \;\;,\;\; (0, - \frac{1}{2}, 0, + \frac{1}{2}) \label{lowering}
\ee
and their Hermitian conjugates
\be
(+ \frac{1}{2}, 0, +\frac{1}{2}, 0) \;\;,\;\; (+ \frac{1}{2}, 0, - \frac{1}{2}, 0) \;\;,\;\; (0, +\frac{1}{2}, 0, + \frac{1}{2}) \;\;,\;\; (0, + \frac{1}{2}, 0, - \frac{1}{2}) . \label{raising}
\ee
We take the components (\ref{lowering}) as lowering operators that act on a highest weight state annihilated by the Hermitian conjugates (\ref{raising}). Taking this highest weight state to have the quantum numbers $(+ \frac{1}{2}, + \frac{1}{2}, 0, 0)$, we obtain states that fill out the representation
\be
(\frac{1}{2}, \frac{1}{2}, 0, 0) \oplus (0, \frac{1}{2}, 0, \frac{1}{2}, 0) \oplus (\frac{1}{2}, 0, 0, \frac{1}{2}) \oplus (0, 0, \frac{1}{2}, \frac{1}{2}) .
\ee
of the group (\ref{group}). The first term is a vector under $SO (4)_L$ and a scalar under $SO (4)_R$. The second (third) term is a Weyl spinor (anti Weyl spinor) under both $SO (4)_L$ and $SO (4)_R$, and the last term is a scalar under $SO (4)_L$ and a vector under $SO (4)_R$. Upon compactification on a circle in the $5$-direction, we obtain precisely a massive vector multiplet in five dimensions.

It is instructive to compare with the case of a massless representation. This means that the charges $z_\mu^a$ and $w_{\mu \nu \rho}^{a b}$ must vanish, and by a Lorentz transformation, we may take the momentum $p_\mu$ to be of the standard form
\be
p_\mu^o = (E, 0, 0, 0, 0, E) 
\ee
for some arbitrarily chosen energy $E$. These data are left invariant by a little group 
\be
SO (4)_L \times SO (5)_R \simeq SU (2) \times SU (2) \times SO (5) , \label{group2}
\ee
under which the supercharge transforms as
\be
(0, \frac{1}{2}, {\bf 4}) \oplus (\frac{1}{2}, 0, {\bf 4}) .
\ee
(Again, we denote $SU (2)$ representations by their spin $j$, whereas $SO (5)$ representations are denoted by their dimensionality in boldface.) In this standard frame, the supersymmetry algebra reads
\be
\left\{ Q_\alpha^i, (Q_\beta^j)^\dagger \right\} = 2 P_{\alpha j}^{\beta i} E ,
\ee
where the projection operator $P_{\alpha j}^{\beta i}$ is given by
\be
P_{\alpha j}^{\beta i} = \frac{1}{2} \delta_j^i \left(\delta_\alpha^\beta + (\Gamma^0 \Gamma^5)_\alpha^\beta \right) .
\ee
The part of the supercharge that survive the projection transform as $(\frac{1}{2}, 0, {\bf 4})$, i.e. we get the components
\be
(-\frac{1}{2}, 0, {\bf 4}) \label{lowering2}
\ee
and their Hermitian conjugates
\be
(+\frac{1}{2}, 0, {\bf 4}) .
\ee
(Here we have labeled the states by their $SU (2)$ $J^z$ eigenvalue $m$ and the dimensionality of the $SO (5)_R$ representation in boldface.) Taking the components (\ref{lowering2}) as lowering operators acting on a highest weight state with quantum numbers $(1, 0, {\bf 1})$, we obtain states filling out the representations
\be
(1, 0, {\bf 1}) \oplus (\frac{1}{2}, 0, {\bf 4}) \oplus (0, 0, {\bf 5}) .
\ee
of the group (\ref{group2}). The first term is a self-dual tensor under $SO (4)_L$ and a singlet under $SO (5)_R$, the second term is a Weyl spinor under $SO (4)_L$ and a spinor under $SO (5)_R$, and the last term is a scalar under $SO (4)_L$ and a vector under $SO (5)_R$, i.e. we get precisely the states of a tensor multiplet.

\section{The quantum fields}
We now return to our massive representation with the states $\left| p, v, \phi, \sigma \right>$. The little group $SO (4)_L \times SO (4)_R$ acts on the $\sigma$ label according to a representation with four irreducible components; an $SO (4)_L$ vector $SO (4)_R$ scalar, an $SO (4)_L$ and $SO (4)_R$ Weyl spinor, an $SO (4)_L$ and $SO (4)_R$ anti Weyl spinor, and an $SO (4)_L$ scalar $SO (4)_R$ vector. We think of these states as being created by applying some operators $a^\dagger (p, v, \phi, \sigma)$ to a vacuum state $\left| 0 \right>$ invariant under the full Lorentz and $R$-symmetry groups $SO (5, 1)$ and $SO (5)$. Our object is now to construct a set of quantum fields for these representations as linear combinations of the creation operators $a^\dagger (p, v, \phi, \sigma)$ and their Hermitian conjugates the annihilation operators $a (p, v, \phi, \sigma)$, with the coefficients chosen so that the fields transform covariantly under $SO (5, 1) \times SO (5)$.

Beginning with the $SO (4)_L$ vector $SO (4)_R$ scalar, the corresponding field is a Lorentz vector $R$-symmetry scalar:
\be
A_\mu (p, v, \phi) = \sum_\sigma e_\mu (p, v, \sigma) a^\dagger (p, v, \phi, \sigma) + {\rm Hermitian \; conjugate}.
\ee
This automatically transforms correctly under the $R$-symmetry. (For the other fields, which are not $R$-symmetry scalars, the coefficient functions will also depend on $\phi$.) Under an arbitrary Lorentz transformation $\Lambda$, the creation operators transform as
\be
a^\dagger (p, v, \phi, \sigma) \rightarrow \frac{N (p, v)}{N (\Lambda p, \Lambda v)} \sum_{\sigma^\prime} W_{\sigma \sigma^\prime} (\Lambda, p, v) a^\dagger (\Lambda p, \Lambda v, \phi, \sigma^\prime) ,
\ee
and the field should transform as
\be
A_\mu (p, v, \phi) \rightarrow \Lambda_\mu{}^\nu A_\nu (\Lambda p, \Lambda v, \phi) ,
\ee
where $\Lambda_\mu{}^\nu$ is the representation matrix of $\Lambda$ in the vector representation. The coefficient functions must thus fulfil the conditions
\be
\Lambda_\mu{}^\nu e_\nu (\Lambda p, \Lambda v, \sigma^\prime) = \frac{N (p, v)}{N (\Lambda p, \Lambda v)} \sum_\sigma e_\mu (p, v, \sigma) W_{\sigma \sigma^\prime} (\Lambda, p, v) . \label{conditions}
\ee
We now take $p$ and $v$ to have their standard values $p^o$ and $v^o$ respectively, and take $\Lambda$ to be the standard Lorentz transformation $L (p, v)$ that transforms these to some arbitrary values $p$ and $v$. Inserting this in (\ref{conditions}) gives a relationship that expresses the coefficient functions $e_\mu (p, v, \sigma)$ for arbitrary $p$ and $v$ in terms of their values at $p^o$ and $v^o$:
\be
e_\nu (p, v, \sigma) = \frac{N (p^o, v^o)}{N (p, v)} L^{-1} (p, v)_\nu{}^\mu e_\mu (p^o, v^o, \sigma) .
\ee
We then take $p$ and $v$ to again equal $p^o$ and $v^o$ and take $\Lambda$ to be an element of the little group $SO (4)_L$ that leaves these invariant. Equation \ref{conditions} then amounts to the following conditions on the coefficient functions $e_\mu (p^o, v^o, \sigma)$:
\be
\Lambda_\mu{}^\nu e_\nu (p^o, v^o, \sigma^\prime) = e_\mu (p^o, v^o, \sigma) \Lambda_{\sigma \sigma^\prime} ,
\ee
where $\Lambda_{\sigma \sigma^\prime}$ is the representation matrix of $\Lambda$ in the vector representation of $SO (4)_L$. This equation can be fulfilled by imposing the polarization conditions
\be
e_0 (p^0, v^0, \sigma) = e_5 (p^0, v^0, \sigma) = 0
\ee
and letting the remaining transverse components of $e_\mu (p^0, v^0, \sigma)$ act as an intertwiner between the $\mu$ and $\sigma$ indices. Rather than working in momentum space, it is of course possible to perform a Fourier transformation to a position dependent field $A_\mu (x, v, \phi)$.

In a rather analogous way, the $SO (4)_L$ scalar $SO (4)_R$ vector gives rise to a field $B^a (x, v, \phi)$ that is a Lorentz scalar $R$-symmetry vector. The $SO (4)_L$ and $SO (4)_R$ spinors lead to fields $\Psi_\alpha^{+ i} (x, v, \phi)$ and $\Psi_\alpha^{- i} (x, v, \phi)$ that are symplectic Majorana spinors under the Lorentz and $R$-symmetry groups and are subject to a certain projection. 

Finally, we will briefly discuss how these fields can be coupled to a chiral two-form $X$ in a tensor multiplet. We then need to construct a two-form current $J$ which is conserved, i.e. $d^* J = 0$. In the presence of such a source, the equation of motion of $X$ reads $d^* H = J$, where $H$ is the field strength of $X$. (By the self-duality of $H$, we also have $d H = * J$.) To construct such a current, we need pairs of charge conjugate fields, e.g. in addition to the field $A_\mu (x, v, \phi)$ we also need its charge conjugate $A^*_\mu (x, v, \phi)$, which simply is another field of the same type. We can then introduce the quantity
\be
{\cal J}_{\mu \nu} (x, v, \phi) = v_{[\mu} \partial_{\nu]} A^\rho (x, v, \phi) A^*_\rho (x, v, \phi) - A^\rho (x, v, \phi)  v_{[\mu} \partial_{\nu]} A^*_\rho (x, v, \phi) ,
\ee
which is conserved since the fields obey the Klein-Gordon equation $(\partial^\mu \partial_\mu - m^2) A^\rho (x, v, \phi) = 0$ and the transversality condition $v^\mu \partial_\mu A^\rho (x, v, \phi) = 0$ and similarly for $A^*_\rho (x, v, \phi)$. (There will also be contributions to ${\cal J}_{\mu \nu} (x, v, \phi)$ from the fields $B^a (x, v, \phi)$, $\Psi^{+ i}_\alpha (x, v, \phi)$ and $\Psi^{- i}_\alpha (x, v, \phi)$ and their charge conjugates.) Integrating this expression over $v$ gives a two-form current $J$ that can be coupled to a chiral two-form. (We remark that this is the analogue of `minimal coupling' in ordinary gauge theory. It has to be modified by further interaction terms to construct a theory that reduces to five-dimensional Yang-Mills theory upon compactification on a circle.)

\vspace*{1mm}
We have benefitted from discussions with Martin Cederwall and Gabriele Ferretti. This research is supported by the Swedish Research Council.

\end{document}